\documentclass[letterpaper]{article} 
\usepackage{aaai25}  
\usepackage{times}  
\usepackage{helvet}  
\usepackage{courier}  
\usepackage[hyphens]{url}  
\usepackage{graphicx} 
\usepackage{natbib}  
\usepackage{caption} 
\usepackage{algorithm}
\usepackage{algorithmic}

\usepackage{newfloat}
\usepackage{listings}
\DeclareCaptionStyle{ruled}{labelfont=normalfont,labelsep=colon,strut=off} 
\floatstyle{ruled}
\newfloat{listing}{tb}{lst}{}
\floatname{listing}{Listing}

\usepackage{xcolor}

\usepackage{longtable}
\usepackage{booktabs}
\usepackage{amsmath}

\title{WikiReddit: Tracing Information and Attention Flows Between Online Platforms
}
\author{
    Patrick Gildersleve\textsuperscript{\rm 1},
    Anna Beers\textsuperscript{\rm 2},
    Viviane Ito\textsuperscript{\rm 2},
    Agustin Orozco\textsuperscript{\rm 2},
    Francesca Tripodi\textsuperscript{\rm 2}
}
\affiliations{
    \textsuperscript{\rm 1} University of Exeter\\
    p.gildersleve@exeter.ac.uk\\
    \textsuperscript{\rm 2} University of North Carolina at Chapel Hill \\
    \{albeers@, itovivi@, aorozco@, ftripodi@email.\}unc.edu

}

\begin{document}
\maketitle

\begin{abstract}

The World Wide Web is a complex interconnected digital ecosystem, where information and attention flow between platforms and communities throughout the globe. These interactions co-construct how we understand the world, reflecting and shaping public discourse.  Unfortunately, researchers often struggle to understand how information circulates and evolves across the web because platform-specific data is often siloed and restricted by linguistic barriers. To address this gap, we present a comprehensive, multilingual dataset capturing all Wikipedia mentions and links shared in posts and comments on Reddit 2020--2023, excluding those from private and NSFW subreddits. Each linked Wikipedia article is enriched with revision history, page view data, article ID, redirects, and Wikidata identifiers. Through a research agreement with Reddit, our dataset ensures user privacy while providing a query and ID mechanism that integrates with the Reddit and Wikipedia APIs. This enables extended analyses for researchers studying how information flows across platforms. For example, Reddit discussions use Wikipedia for deliberation and fact-checking which subsequently influences Wikipedia content, by driving traffic to articles or inspiring edits. By analyzing the relationship between information shared and discussed on these platforms, our dataset provides a foundation for examining the interplay between social media discourse and collaborative knowledge consumption and production.

\end{abstract}

%
\begin{links}
    \link{Code}{https://github.com/pgilders/WikiReddit}
    \link{Dataset}{https://doi.org/10.5281/zenodo.14653265}
\end{links}

\section{Introduction}

Researching flows of information, attention, and discussion on online platforms, and how they both reflect and shape public discourse, is a key challenge in understanding the dynamics of the modern web. There is ample work studying these effects within individual communities and platforms \cite{crane2008robust,lehmann2012dynamical,twyman_black_2017,kobayashi2021modeling,shen_tale_2022,johnson_global_2021,matsui_throw_2024}. While this foundational work set the stage for the importance of computational social science, the web is a complex, interconnected ecosystem. Users freely move between platforms, communities, and devices. To better understand how information propagates and spreads, we must consider interactions across platforms. Indeed, an increasing number of multi-platform studies have identified effects not observed or observable within individual platforms \cite{mcdowall_sensemaking_2024,kloo_cross-platform_2024,moyer_determining_2015, dubois2018echo}. Unfortunately, undertaking this task in research is becoming increasingly difficult. The deterioration of the open web towards that of large centralized platforms where data is often walled off behind APIs, paywalls, or other restrictions, has made it harder to study the web as a whole \citep{freelon_computational_2018,de_vreese_data_2023}. Research opportunities that can bridge these silos are thus of great value.

Wikipedia, as the globally recognized repository of collective knowledge, and Reddit, the foremost social news aggregation and forum site, form an intriguing pairing in this ecosystem. Links to Wikipedia articles on Reddit often serve as references, fact-checking tools, or catalysts for discussion, making them a key touchpoint for understanding how collective attention and knowledge circulates and evolves online \cite{moyer_determining_2015}. However, foundational work on this relationship documented a ``paradox of re-use;'' private platforms like Reddit boost engagement and revenue by relying on Wikipedia links but the relationship is not reciprocal \cite{taraborelli_sum_2015,vincent_examining_2018}. These studies documented the uneven value derived from Wikipedia volunteerism and shed light on how Wikipedia data sustains the economic interests of private platforms and satisfies informational needs.  By expanding on this important work, our database will help researchers better understand the social and information dynamics of these interactions. How Wikipedia links are shared, interpreted, and acted upon within Reddit communities and the wider web---remains relatively underexplored.

We present WikiReddit, a comprehensive dataset capturing all Wikipedia mentions (including links) shared in posts and comments on Reddit from 2020 to 2023, excluding those from private and NSFW (not safe for work) subreddits. The SQL database comprises 336K total posts, 10.2M comments, 1.95M unique links, and 1.26M unique articles spanning  59 languages on Reddit and 276 Wikipedia language subdomains. Each linked Wikipedia article is enriched with its revision history and page view data within a ±10-day window of its posting, as well as article ID, redirects, and Wikidata identifiers. Supplementary anonymous metadata from Reddit posts and comments further contextualizes the links, offering a robust resource for analysing cross-platform information flows, collective attention dynamics, and the role of Wikipedia in online discourse. This dataset is distinct from those used in other works documenting similar relationships between Wikipedia and Reddit due to its scope (four years), diverse supporting data included, and approach to providing a long-term, sustainable resource, using officially licensed Reddit4Researchers API access. This approach ensures the long-term availability of the resource while adhering to ethical standards and user expectations around data use.  By ensuring sustainability, our dataset not only allows for initial analysis in this paper and beyond, but also creates an opportunity for longitudinal analysis moving forward.

In initial explorations of our dataset, we study the use and success of Wikipedia links on Reddit over time, the association between Wikipedia articles being posted on Reddit and changes in page view and editing activity, and finally the distribution of languages used in Reddit posts and linked to on Wikipedia. We find Wikipedia is being referred to in Reddit posts less frequently over the period of study, but the performance of those posts remains stable. There are notable associations between activity on Reddit and Wikipedia; article page views tend to increase on the day the link posts to Reddit and continue, less markedly for a week after. However, the relationship is much weaker for editing activity. Regarding languages, English is the dominant language in the Reddit data, and this is reflected in the Wikipedia links posted. However, there is substantial cross-lingual linking to and from English.

In the following sections, we review related work, articulate how the WikiReddit dataset was developed and summarize its structure. We then undertake some exploratory analysis, before concluding with suggested applications and future research that may be undertaken with the dataset.

\section{Related Work}

Many studies have documented the important role Wikipedia plays in how people find and validate information. Wikipedia content appears in over 80\% of knowledge panels and top-linked content across three different search engines (Google, Bing, and DuckDuckGo) comprising a large portion of most user-facing knowledge graph assets \cite{mcmahon_substantial_2017,vincent_deeper_2021,vincent_examining_2018}. Given the tight integration between search engines and Wikipedia, these studies shed light on how Wikipedia impacts human decision-making and influences other knowledge classification systems \cite{lerner_knowledge_2018,c_thompson_user-generated_2024,formisano_counter-misinformation_2024}. Corporate-owned platforms also depend on Wikipedia data. Many websites use Wikipedia hyperlinks and content to increase visitation, engagement, and revenue \cite{gomezmartinez_wikipedia_2022,lerner_knowledge_2018,moyer_determining_2015,vincent_examining_2018}.  Unfortunately, this economic dependency appears nonreciprocal---while Wikipedia’s open licenses make it easy for corporate sites to capitalize on its content, it does not produce migratory benefits like more viewership or edits on Wikipedia itself \cite{vincent_examining_2018}. 

In addition to Wikipedia, audiences consult a range of online services for news including news websites, apps, and social media \cite{st_aubin_news_2024,vraga_news_2021}. People who include, but do not exclusively rely on, social media in their news diets tend to have higher news media knowledge \cite{schulz_role_2024}. Among these social media sites is Reddit, a social media platform where community members share, vote, and comment on content and foster community-based engagement on topics or themes---colloquially referred to as subreddits. Previous research has explored the relationship between news coverage and Reddit engagement. \citet{gozzi_collective_2020} demonstrated that COVID-19 news coverage drove users to comment on Reddit and search for information on Wikipedia, though this effect decreased over time---probably due to media saturation. Further research on Reddit has found that fact-checked information lasts longer when a post is deemed true \cite{bond_engagement_2023}, reinforcing that users rely on Reddit's discussions when interpreting news. However, other work has scrutinized the credibility of information posted to Reddit---without editorial oversight, users may share biased viewpoints and content from known misinformation sources \cite{chipidza_ideological_2022}. Tangled into these findings on social media and news consumption is the role Wikipedia might still play in this process. Studies have found that Reddit users regularly rely on Wikipedia hyperlinks to validate information---especially within the ``Today I Learned'' subreddit \cite{moyer_determining_2015, vincent_examining_2018}. Nonetheless, access to this previous dataset is no longer feasible---since 2023 Reddit’s API is no longer available for free public use. Researchers wishing to access Reddit data must now submit an application for access to the ``Reddit4Researchers'' beta program---a new approach to partnering with data scientists to balance between data accessibility and user protection \cite{perez_reddit_2024}.

Datasets like the one we present in this paper are part of a long line of research committed to making Wikipedia data accessible and available to other social scientists. Previous papers have built Wikipedia datasets to assess the quality of content on Wikipedia \cite{das_language-agnostic_2024}; study its hyperlink structure \cite{consonni_wikilinkgraphs_2019}; understand how people interact with ‘news events’ \cite{gildersleve_between_2023}; and document platform interdependencies \cite{meier_twikiltwitter_2022}. The organizational structure of the site, its size, and the fact that Wikipedia is open access facilitate these dataset creations \cite{mitrevski_wikihist_2020}. These studies are also pushing the boundaries of Anglocentrism, creating databases that leverage ``language-agnostic'' or multilingual techniques to identify linguistic gaps, explain the informational needs of marginalized populations, and analyze how ideas propagate across the languages \cite{das_language-agnostic_2024,valentim_tracking_2021,miquel-ribe_wikipedia_2019}. Creating these datasets is no simple task, Wikipedia is a massive corpus of densely interlinked content, not just a ``ready-made data source'' \cite{gildersleve_between_2023,valentim_tracking_2021}. 


\section{Dataset Development}

\subsection{Overview}

The dataset is shared as a SQLite3 database via Zenodo:  \url{https://doi.org/10.5281/zenodo.14653265}. Replication code for collection, exploratory analysis, and demo code is provided in the project repository: \url{https://github.com/pgilders/WikiReddit}. Data is collected from the Reddit4Researchers and Wikipedia APIs \cite{a2024_apimain}. This data from these APIs is licensed under the ``Reddit License'' \cite{_2024_developer} and CC BY-SA 4.0 respectively. Following the drastic changes made to the old Reddit API, as well as the shutdown of secondary tools and archives such as Pushshift \citep{mehta_2024_social}, Reddit has opened a new API for researchers program, which this project relies on and aims to integrate into for future data gathering \cite{a2024_our}. We used the WikiToolkit \citep{patrickgildersleve_2023_pgilderswikitoolkit} Python package for fast, reliable collection for a variety of Wikipedia data from their APIs and dumps. Data collection and demo code using WikiToolkit is provided in the repository for this article.

\subsection{Reddit Data}

We used API access to the Reddit4Researchers program to collect data from Reddit. Data is available for 4 years (2020-2023). We collected all posts that mention Wikipedia, either in the title or post body, including content and associated metadata. We also collected all comments on Reddit that mention Wikipedia including content and associated metadata.

For the purposes of this dataset, only post, comment, and subreddit IDs, together with anonymous metadata such as timestamps, score, and extracted Wikipedia links, are shared. All IDs are securely hashed using SHA-256, and checked for uniqueness. This measure preserves individuals' privacy and also enables future researchers to collect and analyze additional data on entries of interest with access to the Reddit4Researchers API by matching against our hashes. This data is stored in the \texttt{posts} and \texttt{comments} tables.

\subsection{Parsing URLs to Articles}

Not every post/comment mentioning Wikipedia includes a Wikipedia URL, and not all posted Wikipedia URLs are valid or correctly formatted such that they map to a valid Wikipedia page. Furthermore, there are a variety of ways in which a Wikipedia URL can link to an article (e.g., directly, via a redirect, to a specific revision). We developed a strategy to reliably extract these URLs and identify which articles they link to. The procedure is outlined as follows:

\begin{enumerate}
    \item Parse the post / comment text with a markdown parser, and extract all correctly hyperlinked Wikipedia URLs.
    \item For any malformed links from markdown and the full remaining text, run a split and regex for any further Wikipedia links.
    \item For all extracted links, run a validation step to ensure it successfully connects to Wikipedia
    \item If the link does not successfully connect, run cleaning steps with regex, removing unnecessary trailing characters.
    \item Recheck validity, and resolve any http redirects  as necessary for all URLs.
    \item For all links, try to extract the Wikipedia language subdomain and article title with regex.
    \item If the link is indirect (e.g., to a revision ID), query the Wikipedia API to get the article title.
    \item Query the Wikipedia API with the extracted article titles to identify any Wikipedia article redirects. Return this as `canonical\_title'.
\end{enumerate}
 
This data is stored in the \texttt{post\_links} (all links from posts), \texttt{comment\_links} (all links from comments), and \texttt{links\_articles} (unique valid links that map to an article) tables. A summary of all Wikipedia mentions, links, and articles posted to Reddit is provided in Table \ref{tab:linksummary}. 

\begin{table}[h]
\centering
\resizebox{\columnwidth}{!}{%
\begin{tabular}{l|rr|rr}
                                  & Posts & Comments & Total\\ \hline
\# mentioning Wikipedia           &  335,897         &  10,264,340   & 10,600,237 \\
\# with Wikipedia links           &  286,359         &  9,465,316    &  9,751,675 \\
Total \# of Wikipedia links       &  658,493         &  11,573,36    & 12,231,860 \\
\# unique Wikipedia links         &  295,439         &  1,890,497    &  1,954,003 \\
\# unique Wikipedia articles      &  252,846         &  1,196,494    &  1,260,479
\end{tabular}}
\caption{A summary of all Wikipedia mentions, those including links, and those that map to Wikipedia articles by Reddit format.}
\label{tab:linksummary}
\end{table}

\subsection{Wikipedia Data}

Wikipedia data is collected via the Wikipedia APIs using WikiToolkit.

\subsubsection{IDs and Redirects}

Wikipedia links typically link to an article name. We resolved this name, which might be outdated, or non-canonical, to the current canonical title (i.e., resolve redirects), collected the page ID, and collected all pages that redirect to the article canonical title. In cases where links are to a page ID or revision ID, we similarly gathered the appropriate page ID, canonical title, and redirects as appropriate (as previously indicated). These are stored in the \texttt{wiki\_ids}, \texttt{resolved\_redirects}, and \texttt{collected\_redirects} tables.

\subsubsection{Page Views}

Daily page view counts are collected for every article posted to Reddit $\pm10$ days from initially post/comment date (\texttt{created\_at}) and $\pm10$ days from last modified date (\texttt{updated\_at}, \texttt{last\_modified\_at}). Page views are collected for both the original posted page title and any canonical redirected title. These are stored in the \texttt{page\_views} table.

\subsubsection{Revisions}

For every article posted to Reddit all article revisions (IDs and timestamps) made $\pm10$ days from initially post/comment date (\texttt{created\_at}) and $\pm10$ days from last modified date (\texttt{updated\_at}, \texttt{last\_modified\_at}) are collected. In addition, the revision at the start of this time period is collected, regardless of when it was created (i.e., the state of the article -10 days from initial post/comment date). These are stored in the \texttt{revisions} table.


\subsection{Wikidata Data}

For each Wikipedia article collected, we also collected their Wikidata identifier (if present), this is also stored in the \texttt{wiki\_ids} table. This allows for cross-referencing the Reddit and Wikipedia data with Wikidata's structured knowledge graph, as well as interlanguage concept resolution.

\subsection{Summary}

A summary of the database structure is provided in Table \ref{tab:sqldb}.
\begin{table*}[h!]
\centering
\resizebox{0.75\textwidth}{!}{%
\begin{tabular}{lll}
\toprule
\textbf{Column Name} & \textbf{Type} & \textbf{Description} \\
\midrule
\multicolumn{3}{l}{\textbf{Table: \texttt{posts}}} \\
\texttt{subreddit\_id} & TEXT & The unique identifier for the subreddit. \\
\texttt{crosspost\_parent\_id} & TEXT & The ID of the original Reddit post if this post is a crosspost. \\
\texttt{post\_id} & TEXT & Unique identifier for the Reddit post. \\
\texttt{created\_at} & TIMESTAMP & The timestamp when the post was created. \\
\texttt{updated\_at} & TIMESTAMP & The timestamp when the post was last updated. \\
\texttt{language\_code} & TEXT & The language code of the post. \\
\texttt{score} & INTEGER & The score (upvotes minus downvotes) of the post. \\
\texttt{upvote\_ratio} & REAL & The ratio of upvotes to total votes. \\
\texttt{gildings} & INTEGER & Number of awards (gildings) received by the post. \\
\texttt{num\_comments} & INTEGER & Number of comments on the post. \\
\midrule
\multicolumn{3}{l}{\textbf{Table: \texttt{comments}}} \\
\texttt{subreddit\_id} & TEXT & The unique identifier for the subreddit. \\
\texttt{post\_id} & TEXT & The ID of the Reddit post the comment belongs to. \\
\texttt{parent\_id} & TEXT & The ID of the parent comment (if a reply). \\
\texttt{comment\_id} & TEXT & Unique identifier for the comment. \\
\texttt{created\_at} & TIMESTAMP & The timestamp when the comment was created. \\
\texttt{last\_modified\_at} & TIMESTAMP & The timestamp when the comment was last modified. \\
\texttt{score} & INTEGER & The score (upvotes minus downvotes) of the comment. \\
\texttt{upvote\_ratio} & REAL & The ratio of upvotes to total votes for the comment. \\
\texttt{gilded} & INTEGER & Number of awards (gildings) received by the comment. \\
\midrule
\multicolumn{3}{l}{\textbf{Table: \texttt{postlinks}}} \\
\texttt{post\_id} & TEXT & Unique identifier for the Reddit post. \\
\texttt{end\_processed\_valid} & INTEGER & Whether the extracted URL from the post resolves to a valid URL. \\
\texttt{end\_processed\_url} & TEXT & The extracted URL from the Reddit post. \\
\texttt{final\_valid} & INTEGER & Whether the final URL from the post resolves to a valid URL after any redirections. \\
\texttt{final\_status} & INTEGER & HTTP status code of the final URL. \\
\texttt{final\_url} & TEXT & The final URL after any redirections. \\
\texttt{redirected} & INTEGER & Indicator of whether the posted URL was redirected (1) or not (0). \\
\texttt{in\_title} & INTEGER & Indicator of whether the link appears in the post title (1) or post body (0). \\
\midrule
\multicolumn{3}{l}{\textbf{Table: \texttt{commentlinks}}} \\
\texttt{comment\_id} & TEXT & Unique identifier for the Reddit comment. \\
\texttt{end\_processed\_valid} & INTEGER & Whether the extracted URL from the comment resolves to a valid URL. \\
\texttt{end\_processed\_url} & TEXT & The extracted URL from the comment. \\
\texttt{final\_valid} & INTEGER & Whether the final URL from the comment resolves to a valid URL after any redirections. \\
\texttt{final\_status} & INTEGER & HTTP status code of the final URL. \\
\texttt{final\_url} & TEXT & The final URL after any redirections. \\
\texttt{redirected} & INTEGER & Indicator of whether the URL was redirected (1) or not (0). \\
\midrule
\multicolumn{3}{l}{\textbf{Table: \texttt{linkarticles}}} \\
\texttt{final\_url} & TEXT & The final URL after any redirections. \\
\texttt{lang} & TEXT & The language code of the page. \\
\texttt{mobile} & INTEGER & Indicator of whether the link was mobile-specific (1) or not (0). \\
\texttt{raw\_title} & TEXT & The raw, unprocessed title text extracted from the link. \\
\midrule
\multicolumn{3}{l}{\textbf{Table: \texttt{resolved\_redirects}}} \\
\texttt{lang} & TEXT & The language code of the Wikipedia page. \\
\texttt{raw\_title} & TEXT & The raw title of the from the Wikipedia link before redirection. \\
\texttt{norm\_title} & TEXT & The normalized raw title of the page. \\
\texttt{canonical\_title} & TEXT & The canonical title after resolving the redirect. \\
\midrule
\multicolumn{3}{l}{\textbf{Table: \texttt{collected\_redirects}}} \\
\texttt{lang} & TEXT & The language code of the Wikipedia page. \\
\texttt{canonical\_title} & TEXT & The canonical title of the page. \\
\texttt{other\_title} & TEXT & Other titles associated with the page that redirect to the canonical title. \\
\midrule
\multicolumn{3}{l}{\textbf{Table: \texttt{wiki\_ids}}} \\
\texttt{lang} & TEXT & The language code of the Wikipedia page. \\
\texttt{title} & TEXT & The title of the Wikipedia page. \\
\texttt{pageid} & INTEGER & Unique identifier for the page in Wikipedia. \\
\texttt{wikidata\_id} & TEXT & The Wikidata identifier for the page. \\
\midrule
\multicolumn{3}{l}{\textbf{Table: \texttt{pageviews}}} \\
\texttt{lang} & TEXT & The language code of the Wikipedia page. \\
\texttt{title} & TEXT & The title of the Wikipedia page (not strictly the canonical title). \\
\texttt{date} & TIMESTAMP & The date of the page view count. \\
\texttt{pageviews} & INTEGER & The number of page views on the given date. \\
\midrule
\multicolumn{3}{l}{\textbf{Table: \texttt{revisions}}} \\
\texttt{lang} & TEXT & The language code of the Wikipedia page. \\
\texttt{canonical\_title} & TEXT & The canonical title of the Wikipedia page. \\
\texttt{revid} & INTEGER & The unique revision identifier. \\
\texttt{parentid} & INTEGER & The ID of the parent revision. \\
\texttt{timestamp} & TEXT & The timestamp of the revision. \\
\bottomrule
\end{tabular}%
}
\caption{SQL Database Schema with all tables. All IDs from Reddit are hashed using SHA-256 for anonymization.}
\label{tab:sqldb}
\end{table*}

\section{Ethical and FAIR Considerations}

This work was approved by the University of Exeter ethical review procedures (ID 8969882). The dataset is based on public data collected via the Reddit4Researchers and Wikipedia APIs. This research was observational, and no personally identifiable information from Reddit users was included. Data is limited to publicly accessible posts and comments containing Wikipedia links, excluding private and NSFW subreddits. The dataset conforms to the FAIR principles \cite{wilkinson_fair_2016} as follows.
\begin{itemize}
    \item \textbf{Findable}: The dataset is publicly available on Zenodo and is assigned a permanent DOI: 10.5281/zenodo.14653265
    \item \textbf{Accessible}: Anyone with an internet connection can freely access the dataset, which is shared under a licensing agreement that ensures long-term availability and responsible use.
    \item \textbf{Interoperable}: The dataset is provided in SQLite3 format and includes multilingual and cross-platform identifiers to enhance compatibility and facilitate integration with other tools and datasets.
    \item \textbf{Re-usable}: The dataset comes with replication and demo code for data collection, and exploratory analysis, making it easy for researchers to reproduce or extend analyses. The dataset is licensed CC BY 4.0.
\end{itemize}
This dataset could be misused by researchers attempting to make claims about Wikipedia's political effects on other platforms as a basis for erroneously discrediting the Wikipedia platform, which has recently faced increased politicization \citep{rascouet-paz_elon_2024}. We rely on the academic community to audit and resist bad faith usage of this dataset. Users do not explicitly consent to data collection because their posts are made on a public platform---although researchers have noted that some users, nonetheless, prefer to have their work not used for research \citep{fiesler_participant_2018}. Because this dataset online includes IDs linking to usernames, posts, and comments, users can effectively remove their data from future re-collection and use by deleting their associated data on Reddit.

\section{Exploratory Analysis}

\subsubsection{Reddit \& Wikipedia Use Over Time}

\begin{figure}
    \centering
    \includegraphics[width=\linewidth]{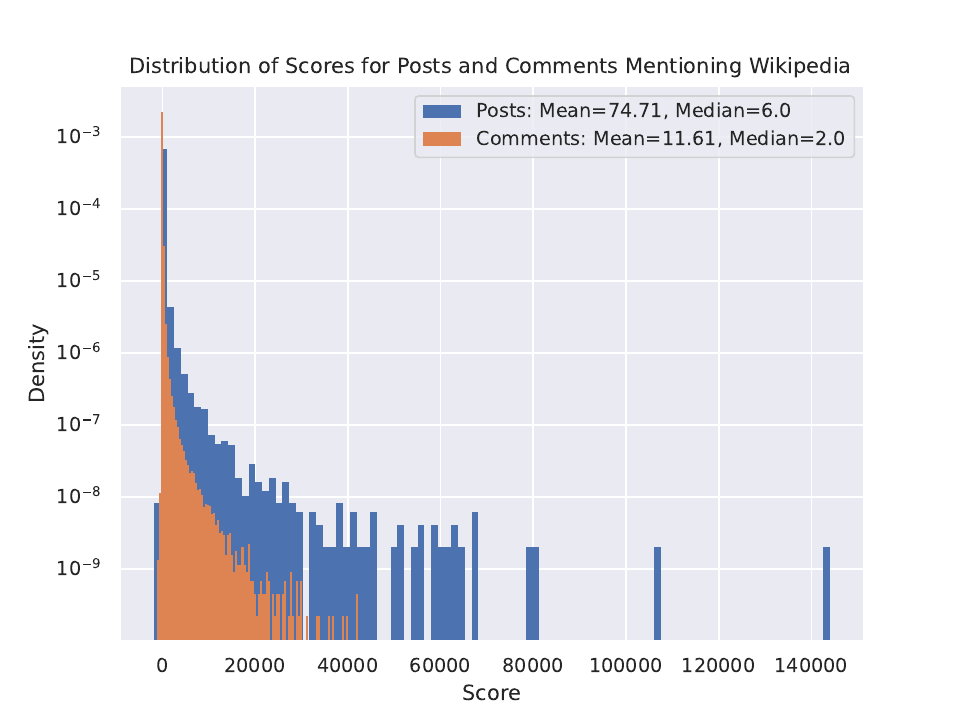}
    \caption{Histograms for the Reddit score of the posts and comments that mention Wikipedia (in text or as a link).}
    \label{fig:posts_comments_score_distribution}
\end{figure}

\begin{figure}
    \centering
    \includegraphics[width=\linewidth]{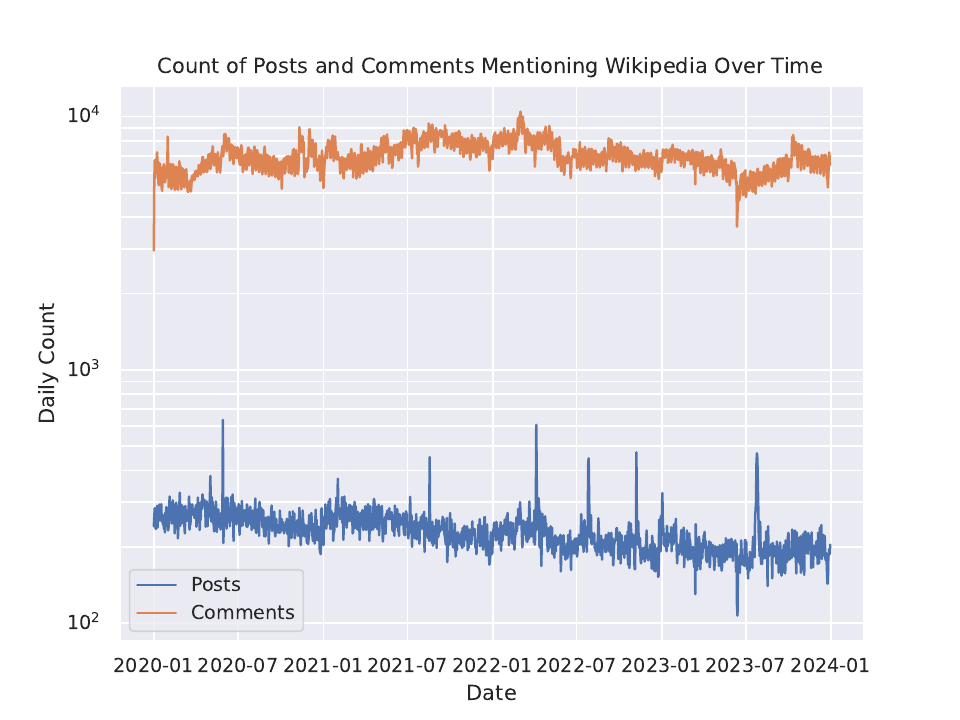}
    \caption{Plot showing the daily count of posts and comments that mention Wikipedia (in text or as a link) over 2020-2023.}
    \label{fig:posts_comments_over_time}
\end{figure}

\begin{figure}
    \centering
    \includegraphics[width=\linewidth]{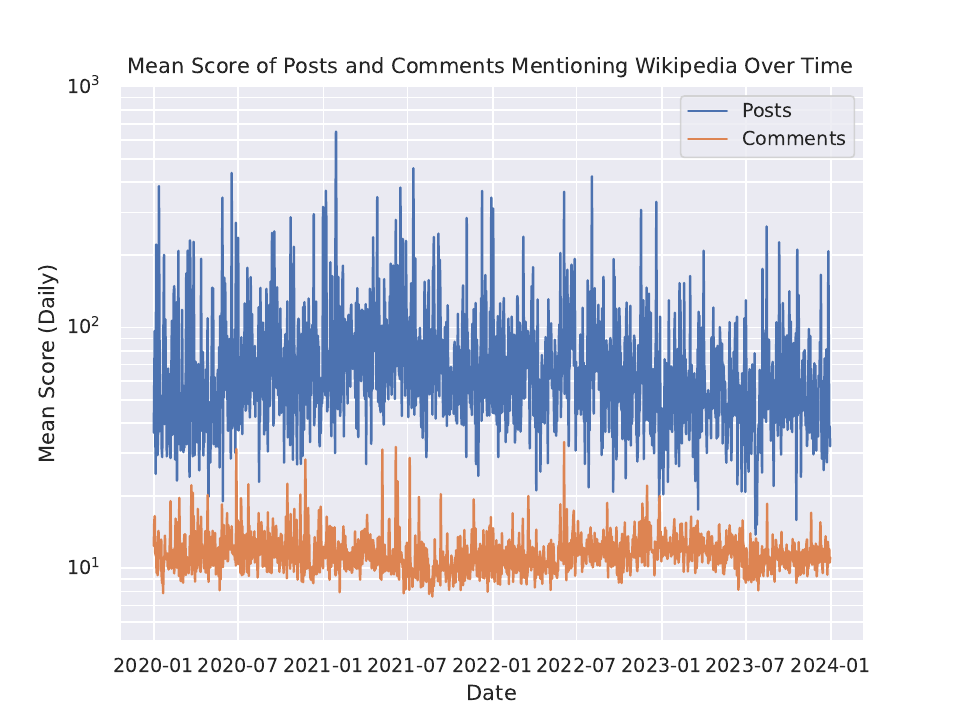}
    \caption{Plot showing the daily average Reddit score of the posts and comments that mention Wikipedia (in text or as a link) over 2020--2023.}
    \label{fig:posts_comments_score_over_time}
\end{figure}

The histograms in Figure \ref{fig:posts_comments_score_distribution} and line plots in Figures \ref{fig:posts_comments_over_time} and \ref{fig:posts_comments_score_over_time} document the distribution of Reddit posts or comments that mention Wikipedia either by name or with an article link over time. The long tail distributions of Fig. \ref{fig:posts_comments_score_distribution} indicates that the vast majority of posts and comments have a score $\approx 0$, whereas a small number prove to be very high scoring. Unsurprisingly, since they are more readily presented to more Reddit users, posts tend to score more highly than comments.

When we analyze the distributions longitudinally, we find that Wikipedia is referenced about 8000 times per day in comments and that number has remained relatively stable over time. Original posts that refer to Wikipedia have decreased slightly over time going from approximately 300 posts per day to around 200 posts per day. While not fully answerable solely with our dataset, we hypothesize that these shifts might be related to changes in platform content moderation, engagement preferences with both platforms (e.g., mobile vs laptop) or other technological affordances (e.g., Reddit’s transition to more image and video-based participation).

\subsubsection{Wikipedia Activity}

Information from Wikipedia being posted to Reddit may be an indicator of what kinds of information people are paying most attention to. Topics of news attention often play out on Wikipedia \cite{gildersleve_between_2023}, but these attention patterns---sometimes together with the Wikipedia articles to help contextualize the events---also move throughout the internet. Much like the \citet{gozzi_collective_2020} findings, our data indicates that people trying to make sense of news as it unfolds go between both platforms. This may be reflected in the Wikipedia page view and editing patterns of the posted articles. The posting of articles on Reddit may itself also drive interest and editing activity towards Wikipedia.

\begin{figure}
    \centering
    \includegraphics[width=\linewidth]{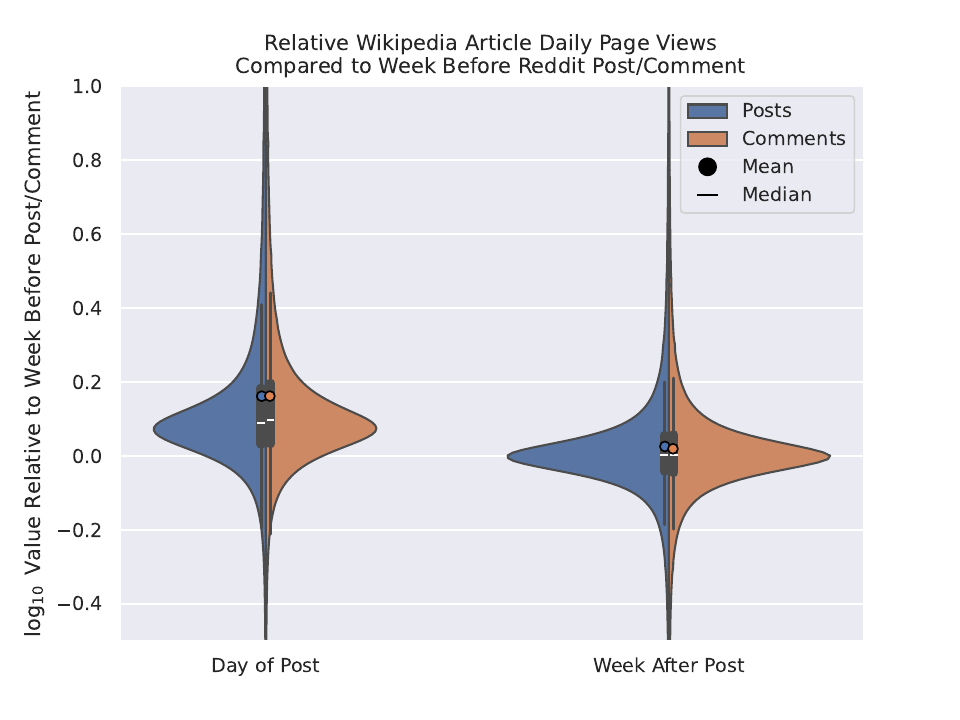}
    \caption{Figure showing the daily page views to Wikipedia articles on the day of posting and in the week after posting relative to the week before posting. A small number of points in the extremes of the distributions are cut for visual clarity.}
    \label{fig:pageview_change_violin}
\end{figure}

\textbf{Page views:} As an indication of these effects, we calculate and plot (Fig. \ref{fig:pageview_change_violin}) the relative values of daily page views the day an article is posted and week after it is posted as compared to the week before it is posted ($\frac{\text{Views on date}}{\text{Mean views in week before}}$ and  $\frac{\text{Mean views in week after}}{\text{Mean views in week before}}$, where the views on the date of posting are not included in either of the other ranges). 

For page views, we first deal with 0 counts, typically indicating an article doesn’t yet exist or is deleted by removing them from analysis. We then compute the geometric mean of the relative values, since the distribution is extremely long-tailed. We find that for links in Reddit posts, we observe a 45\% increase in page views on the day of posting, and a 6\% increase in the week after posting, as compared to the week before posting. For links in Reddit comments, we observe a 45\% increase in page views on the day of posting, and a 5\% increase in the week after posting, as compared to the week before posting.

The similarity between posts and comments in Figure \ref{fig:pageview_change_violin} would suggest much of this activity is due to some external stimulus, as one would assume the level of attention to posts vs comments, and the subsequent spillovers to Wikipedia, would be different. This indicates that audiences are not passive news receivers, but consult different online platforms to support their comprehension of current events. 

\textbf{Edits:} The picture regarding edits is less clear. This is partially due to the fact that many of the linked articles receive no edits in the days before / on / after being posted on Reddit, in line with the findings of \citet{vincent_examining_2018}. Nevertheless, we do see some indications of increased editing activity in the wake of being posted. We first remove cases where the pages receive no edits over the full time period, then, due to the smaller, discrete edit values, consider the absolute change in number of daily edits. We compare the number of daily edits on the day of posting on Reddit and in the week after posting against that in the week before posting. Considering the arithmetic mean, for links in Reddit posts, we observe an increase of 0.462 daily edits on the day of posting, but a decrease of 0.031 daily edits in the week after posting, as compared to the week before posting. For links in Reddit comments, we observe an increase of 0.127 daily edits on the day of posting, but a decrease of 0.022 daily edits in the week after posting, as compared to the week before posting. The relatively small effect size here, plus high levels of zero-inflation, and the presence of extreme outliers warrants more rigorous analysis.

To be clear, we present this analysis as a demonstration of association, rather than an investigation of any causal relationship between Reddit and Wikipedia cross-posting. When there is an association, it is likely in the majority of cases that the Reddit posting and page view / editing behavior is mostly in response to some common external stimuli (see discussion of Fig. \ref{fig:pageview_change_violin}). Studies interrogating any causal relationships are most welcome to be performed using the dataset, making the appropriate decisions for research design, data subsetting, and controls.

\subsubsection{Linking by Language}

\begin{figure}
    \centering
    \includegraphics[width=\linewidth]{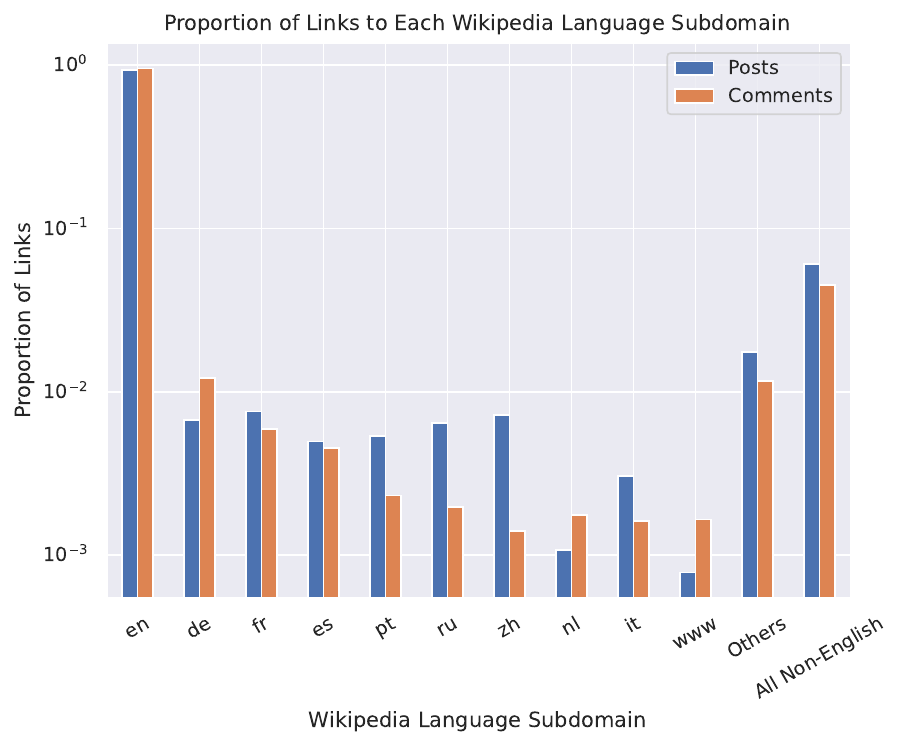}
    \caption{Figure showing the proportion of links to each Wikipedia language subdomain for the most frequently occurring languages in the dataset.}
    \label{fig:lang_proportion}
\end{figure}

\begin{figure}
    \centering
    \includegraphics[width=\linewidth]{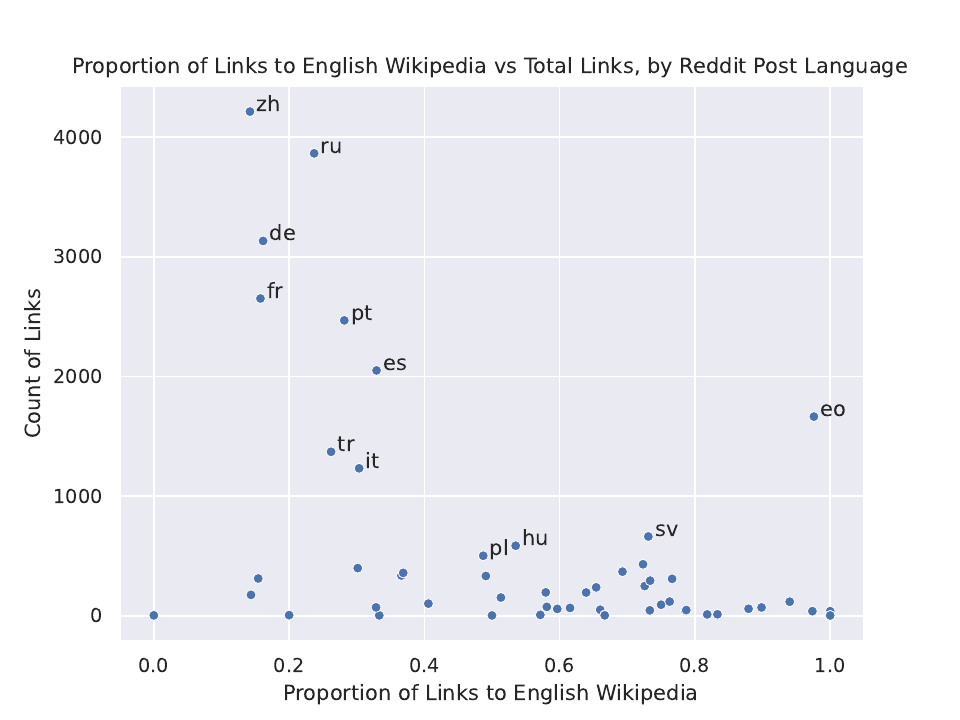}
    \caption{Figure showing the proportion of links to English Wikipedia from non-English Reddit posts vs the total number of links in that post language.}
    \label{fig:en_proportion_vs_total}
\end{figure}

Both Reddit and Wikipedia are used in different languages. Regarding the use of Wikipedia links on the social media platform, activity is very much English language dominated. 95.8\% of Reddit posts with Wikipedia links in the dataset are in English and 93.9\% of all linked articles are to English Wikipedia, with the next most links being to German, French, and Spanish Wikipedias (Fig. \ref{fig:lang_proportion}). However, there is substantial cross-lingual linking. 50.85\% of links to non-English Wikipedias come from English language Reddit posts. It is also notable how frequently non-English language Reddit posts link to the English language Wikipedia (Fig. \ref{fig:en_proportion_vs_total}). The most active other languages are less likely to link to English Wikipedia. An interesting dynamic is observed for Esperanto (eo), the artificially constructed international second language. Esperanto Wikipedia exists, but Esperanto users on Reddit almost exclusively use English Wikipedia—presumably their first language.

The use of non-English Wikipedia articles in English Reddit posts may highlight knowledge gaps on the English Wikipedia, or be used to compare accounts in different language editions. The extensive use of English Wikipedia in other language Reddit posts strongly highlights (perceived) knowledge gaps in these languages. Unfortunately, it is likely that this linking behavior weakens the already fragile pipeline of contributions coming from social media, lessening the likelihood of edits being made to the Wikipedia language editions in need of them. These results, and future work, can have wide-ranging implications for the multilingual state of Wikipedia and the web. 

\section{Conclusion and Future Work}

We present WikiReddit, a comprehensive dataset capturing all Wikipedia mentions (including links) shared in posts and comments on Reddit from 2020 to 2023, excluding those from private and NSFW (not safe for work) subreddits. The SQL database comprises 336K total posts, 10.2M comments, 1.95M unique links, and 1.26M unique articles spanning  59 languages on Reddit and 276 Wikipedia language subdomains. 
Our exploratory analysis reveals Wikipedia is being referred to in Reddit posts less frequently over the period of study, but the performance of those posts remains stable. We also find notable associations between activity on Reddit and Wikipedia; article page views tend to increase on the day the link posts to Reddit and continue, less markedly for a week after. However, the relationship is much weaker for editing activity. Finally, English is unsurprisingly the dominant language in the Reddit posts and comments, which is reflected in the Wikipedia links posted. However, there is substantial cross-lingual linking to and from English.

We foresee several applications of this dataset and preview four here. First, Reddit linking data can be used to understand how attention is driven from one platform to another. Wikipedia, somewhat uniquely among widely-used internet platforms, publishes all of its traffic data for each of its articles on an hourly basis \citep{wikimedia_foundation_page_2024}, facilitating fine-grained analyses of how, for example, Reddit posts and comments direct attention to the encyclopedia. This analysis of cross-platform attention to Wikipedia is particularly worthwhile during times of crisis, where viral traffic driven by social media linking can stress the response capabilities of Wikipedia's volunteer editing community \citep{avieson_breaking_2019,avieson_editors_2022}. 

Second, Reddit linking data can shed light on \textit{how} Wikipedia's archive of knowledge is used in the larger social web. While we have ample research on the uses of Wikipedia within the Wikipedia platform \citep{singer_why_2017}, and tentative research on the on-platform misuses of Wikipedia \citep{saez-trumper_online_2019,kharazian_governance_2024}, we know less about how Wikipedia is used or misused off-platform. Prior research gives us reason to believe that even trustworthy content can be recontextualized outside of its authors' intent, unwittingly supporting disinformation campaigns \citep{beers_selective_2023}. Datasets like these can help us understand how the intellectual work of Wikipedia editors travel, and potentially inform how to make Wikipedia's information resilient to malicious decontextualization. 

Third, our dataset could provide insights into how external attention is topically distributed across Wikipedia. Many have observed biases both who views Wikipedia and how Wikipedia is edited \cite{johnson_global_2021,tripodi_ms_2023,menking_people_2019}. Our dataset can help extend that analysis into the disparities in what types of external communities Wikipedia is used in, and how it is used.

Fourth, a topic analysis of our dataset could reveal how Wikipedia usage on Reddit contributes to societal benefits and harms. Reddit has taken steps in recent years to reform its content moderation to address documentation of vitriolic discourse (see \citet{farrell2019exploring} or \citet{massanari__2017} for earlier examples of sexism/racism). However, both communities still predominantly consist of white men \citep{gilbert_i_2020, menking_people_2019}, a demographic imbalance that may contribute to larger inequities in information \citep{tripodi_ms_2023}. Our dataset could help examine if homogeneity within these groups shapes topic patterns and assess whether these relationships mitigate or amplify problematic engagement online.

There are some limitations to what can be done with this dataset. Some analyses of WikiReddit, such as those that recover the full text of Reddit comments, required continued access to the Reddit4Researchers API platform. While Reddit has shown a positive stance towards data access for external researchers via is Reddit4Researchers program, we must be cautious in what has been dubbed the ``post-API age'' of limited data access from private companies \citep{freelon_computational_2018}. Researchers should also note that URL-sharing is not the only method by which Wikipedia content is shared across platforms like Reddit. For example, text screenshots of external content, previously implicated in the spreading is misinformation \citep{matatov_stop_2022}, will not be captured here, and surely represent some of the external representation of Wikipedia on Reddit. We thus caution that any full census of Wikipedia's external usage is partial if accessed solely through the in-text mentions and URL-sharing included in this dataset. Nevertheless, the WikiReddit dataset represents an important step forward in understanding how Wikipedia is referenced and engaged with across a major social media platform, and provides a reliable resource for long-term study.

\bibliography{aaai25}

\begin{thebibliography}{56}
\providecommand{\natexlab}[1]{#1}

\bibitem[{Avieson(2019)}]{avieson_breaking_2019}
Avieson, B. 2019.
\newblock Breaking news on {Wikipedia}: collaborating, collating and competing.
\newblock \emph{First Monday}.

\bibitem[{Avieson(2022)}]{avieson_editors_2022}
Avieson, B. 2022.
\newblock Editors, sources and the 'go back' button: {Wikipedia}'s framework for beating misinformation.
\newblock \emph{First Monday}.

\bibitem[{Beers et~al.(2023)Beers, Nguy\'{e}\~{e}n, Starbird, West, and Spiro}]{beers_selective_2023}
Beers, A.; Nguy\'{e}\~{e}n, S.; Starbird, K.; West, J.~D.; and Spiro, E.~S. 2023.
\newblock Selective and deceptive citation in the construction of dueling consensuses.
\newblock \emph{Science Advances}, 9(38): eadh1933.
\newblock Publisher: American Association for the Advancement of Science.

\bibitem[{Bond and Garrett(2023)}]{bond_engagement_2023}
Bond, R.~M.; and Garrett, R.~K. 2023.
\newblock Engagement with fact-checked posts on {Reddit}.
\newblock \emph{PNAS Nexus}, 2(3): pgad018.

\bibitem[{C.~Thompson et~al.(2024)C.~Thompson, Luo, McKenzie, Richardson, and Flanagan}]{c_thompson_user-generated_2024}
C.~Thompson, N.; Luo, X.; McKenzie, B.; Richardson, E.; and Flanagan, B. 2024.
\newblock User-{Generated} {Content} {Shapes} {Judicial} {Reasoning}: {Evidence} from a {Randomized} {Control} {Trial} on {Wikipedia}.
\newblock \emph{Information Systems Research}, 35(4): 1948--1964.

\bibitem[{Chipidza et~al.(2022)Chipidza, Krewson, Gatto, Akbaripourdibazar, and Gwanzura}]{chipidza_ideological_2022}
Chipidza, W.; Krewson, C.; Gatto, N.; Akbaripourdibazar, E.; and Gwanzura, T. 2022.
\newblock Ideological variation in preferred content and source credibility on {Reddit} during the {COVID}-19 pandemic.
\newblock \emph{Big Data \& Society}, 9(1): 20539517221076486.
\newblock Publisher: SAGE Publications Ltd.

\bibitem[{Consonni, Laniado, and Montresor(2019)}]{consonni_wikilinkgraphs_2019}
Consonni, C.; Laniado, D.; and Montresor, A. 2019.
\newblock {WikiLinkGraphs}: a complete, longitudinal and multi-language dataset of the {Wikipedia} link networks.
\newblock In \emph{Proceedings of the {International} {AAAI} {Conference} on {Web} and {Social} {Media}}, volume~13, 598--607.

\bibitem[{Crane and Sornette(2008)}]{crane2008robust}
Crane, R.; and Sornette, D. 2008.
\newblock Robust dynamic classes revealed by measuring the response function of a social system.
\newblock \emph{Proceedings of the National Academy of Sciences}, 105(41): 15649--15653.

\bibitem[{Das et~al.(2024)Das, Johnson, Saez-Trumper, and Aragón}]{das_language-agnostic_2024}
Das, P.; Johnson, I.; Saez-Trumper, D.; and Aragón, P. 2024.
\newblock Language-{Agnostic} {Modeling} of {Wikipedia} {Articles} for {Content} {Quality} {Assessment} across {Languages}.
\newblock In \emph{Proceedings of the {International} {AAAI} {Conference} on {Web} and {Social} {Media}}, volume~18, 1924--1934.

\bibitem[{de~Vreese and Tromble(2023)}]{de_vreese_data_2023}
de~Vreese, C.; and Tromble, R. 2023.
\newblock The {Data} {Abyss}: {How} {Lack} of {Data} {Access} {Leaves} {Research} and {Society} in the {Dark}.
\newblock \emph{Political Communication}, 40(3): 356--360.
\newblock Publisher: Routledge \_eprint: https://doi.org/10.1080/10584609.2023.2207488.

\bibitem[{Dubois and Blank(2018)}]{dubois2018echo}
Dubois, E.; and Blank, G. 2018.
\newblock The echo chamber is overstated: the moderating effect of political interest and diverse media.
\newblock \emph{Information, communication \& society}, 21(5): 729--745.

\bibitem[{Farrell et~al.(2019)Farrell, Fernandez, Novotny, and Alani}]{farrell2019exploring}
Farrell, T.; Fernandez, M.; Novotny, J.; and Alani, H. 2019.
\newblock Exploring misogyny across the manosphere in reddit.
\newblock In \emph{Proceedings of the 10th ACM conference on web science}, 87--96.

\bibitem[{Fiesler and Proferes(2018)}]{fiesler_participant_2018}
Fiesler, C.; and Proferes, N. 2018.
\newblock “{Participant}” {Perceptions} of {Twitter} {Research} {Ethics}.
\newblock \emph{Social Media + Society}, 4(1): 2056305118763366.
\newblock Publisher: SAGE Publications Ltd.

\bibitem[{Formisano et~al.(2024)Formisano, Hine, Juneja, Laitila, Novelli, Chiu, Dejanikus, Levin, Schroder, West, and Floridi}]{formisano_counter-misinformation_2024}
Formisano, G.; Hine, E.; Juneja, P.; Laitila, J.; Novelli, C.; Chiu, E.; Dejanikus, E.; Levin, M.; Schroder, T.; West, A.; and Floridi, L. 2024.
\newblock Counter-{Misinformation} {Dynamics}: {The} {Case} of {Wikipedia} {Editing} {Communities} during the 2024 {US} {Presidential} {Elections}.
\newblock \url{https://papers.ssrn.com/abstract=4990973}.
\newblock Accessed: 2025-01-14.

\bibitem[{Freelon(2018)}]{freelon_computational_2018}
Freelon, D. 2018.
\newblock Computational {Research} in the {Post}-{API} {Age}.
\newblock \emph{Political Communication}, 35(4): 665--668.
\newblock Publisher: Routledge \_eprint: https://doi.org/10.1080/10584609.2018.1477506.

\bibitem[{Gilbert(2020)}]{gilbert_i_2020}
Gilbert, S.~A. 2020.
\newblock "{I} run the world's largest historical outreach project and it's on a cesspool of a website." {Moderating} a {Public} {Scholarship} {Site} on {Reddit}: {A} {Case} {Study} of r/{AskHistorians}.
\newblock \emph{Proc. ACM Hum.-Comput. Interact.}, 4(CSCW1): 19:1--19:27.

\bibitem[{Gildersleve(2023)}]{patrickgildersleve_2023_pgilderswikitoolkit}
Gildersleve, P. 2023.
\newblock pgilders/WikiToolkit.
\newblock \url{https://github.com/pgilders/WikiToolkit}.
\newblock Accessed: 2025-01-15.

\bibitem[{Gildersleve, Lambiotte, and Yasseri(2023)}]{gildersleve_between_2023}
Gildersleve, P.; Lambiotte, R.; and Yasseri, T. 2023.
\newblock Between news and history: identifying networked topics of collective attention on {Wikipedia}.
\newblock \emph{Journal of Computational Social Science}, 6(2): 845--875.

\bibitem[{Gozzi et~al.(2020)Gozzi, Tizzani, Starnini, Ciulla, Paolotti, Panisson, and Perra}]{gozzi_collective_2020}
Gozzi, N.; Tizzani, M.; Starnini, M.; Ciulla, F.; Paolotti, D.; Panisson, A.; and Perra, N. 2020.
\newblock Collective {Response} to {Media} {Coverage} of the {COVID}-19 {Pandemic} on {Reddit} and {Wikipedia}: {Mixed}-{Methods} {Analysis}.
\newblock \emph{Journal of Medical Internet Research}, 22(10): e21597.

\bibitem[{Gómez‐Martínez, Orden‐Cruz, and Martínez‐Navalón(2022)}]{gomezmartinez_wikipedia_2022}
Gómez‐Martínez, R.; Orden‐Cruz, C.; and Martínez‐Navalón, J.~G. 2022.
\newblock Wikipedia pageviews as investors’ attention indicator for {Nasdaq}.
\newblock \emph{Intelligent Systems in Accounting, Finance and Management}, 29(1): 41--49.

\bibitem[{Johnson et~al.(2021)Johnson, Lemmerich, Sáez-Trumper, West, Strohmaier, and Zia}]{johnson_global_2021}
Johnson, I.; Lemmerich, F.; Sáez-Trumper, D.; West, R.; Strohmaier, M.; and Zia, L. 2021.
\newblock Global {Gender} {Differences} in {Wikipedia} {Readership}.
\newblock \emph{Proceedings of the International AAAI Conference on Web and Social Media}, 15: 254--265.

\bibitem[{Kharazian, Starbird, and Hill(2024)}]{kharazian_governance_2024}
Kharazian, Z.; Starbird, K.; and Hill, B.~M. 2024.
\newblock Governance {Capture} in a {Self}-{Governing} {Community}: {A} {Qualitative} {Comparison} of the {Croatian}, {Serbian}, {Bosnian}, and {Serbo}-{Croatian} {Wikipedias}.
\newblock \emph{Proceedings of the ACM on Human-Computer Interaction}, 8(CSCW1): 1--26.

\bibitem[{Kloo, Cruickshank, and Carley(2024)}]{kloo_cross-platform_2024}
Kloo, I.; Cruickshank, I.~J.; and Carley, K.~M. 2024.
\newblock A {Cross}-{Platform} {Topic} {Analysis} of the {Nazi} {Narrative} on {Twitter} and {Telegram} during the 2022 {Russian} {Invasion} of {Ukraine}.
\newblock \emph{Proceedings of the International AAAI Conference on Web and Social Media}, 18: 839--850.

\bibitem[{Kobayashi et~al.(2021)Kobayashi, Gildersleve, Uno, and Lambiotte}]{kobayashi2021modeling}
Kobayashi, R.; Gildersleve, P.; Uno, T.; and Lambiotte, R. 2021.
\newblock Modeling collective anticipation and response on Wikipedia.
\newblock In \emph{Proceedings of the international AAAI conference on web and social media}, volume~15, 315--326.

\bibitem[{Lehmann et~al.(2012)Lehmann, Gon{\c{c}}alves, Ramasco, and Cattuto}]{lehmann2012dynamical}
Lehmann, J.; Gon{\c{c}}alves, B.; Ramasco, J.~J.; and Cattuto, C. 2012.
\newblock Dynamical classes of collective attention in twitter.
\newblock In \emph{Proceedings of the 21st international conference on World Wide Web}, 251--260.

\bibitem[{Lerner and Lomi(2018)}]{lerner_knowledge_2018}
Lerner, J.; and Lomi, A. 2018.
\newblock Knowledge categorization affects popularity and quality of {Wikipedia} articles.
\newblock \emph{PLOS ONE}, 13(1): e0190674.
\newblock Publisher: Public Library of Science.

\bibitem[{Massanari(2017)}]{massanari__2017}
Massanari, A. 2017.
\newblock \# {Gamergate} and {The} {Fappening}: {How} {Reddit}’s algorithm, governance, and culture support toxic technocultures.
\newblock \emph{New media \& society}, 19(3): 329--346.
\newblock Publisher: Sage Publications Sage UK: London, England.

\bibitem[{Matatov, Naaman, and Amir(2022)}]{matatov_stop_2022}
Matatov, H.; Naaman, M.; and Amir, O. 2022.
\newblock Stop the [{Image}] {Steal}: {The} {Role} and {Dynamics} of {Visual} {Content} in the 2020 {U}.{S}. {Election} {Misinformation} {Campaign}.
\newblock \emph{Proc. ACM Hum.-Comput. Interact.}, 6(CSCW2): 541:1--541:24.

\bibitem[{Matsui, Miyazaki, and Murayama(2024)}]{matsui_throw_2024}
Matsui, A.; Miyazaki, K.; and Murayama, T. 2024.
\newblock Throw {Your} {Hat} in the {Ring} (of {Wikipedia}): {Exploring} {Urban}-{Rural} {Disparities} in {Local} {Politicians}' {Information} {Supply}.
\newblock \emph{Proceedings of the International AAAI Conference on Web and Social Media}, 18: 1027--1040.

\bibitem[{McDowall, Antoniak, and Mimno(2024)}]{mcdowall_sensemaking_2024}
McDowall, L.; Antoniak, M.; and Mimno, D. 2024.
\newblock Sensemaking about {Contraceptive} {Methods} across {Online} {Platforms}.
\newblock \emph{Proceedings of the International AAAI Conference on Web and Social Media}, 18: 1041--1053.

\bibitem[{McMahon, Johnson, and Hecht(2017)}]{mcmahon_substantial_2017}
McMahon, C.; Johnson, I.; and Hecht, B. 2017.
\newblock The substantial interdependence of {Wikipedia} and {Google}: {A} case study on the relationship between peer production communities and information technologies.
\newblock In \emph{Proceedings of the {International} {AAAI} {Conference} on {Web} and {Social} {Media}}, volume~11, 142--151.
\newblock Issue: 1.

\bibitem[{MediaWiki(2024)}]{a2024_apimain}
MediaWiki. 2024.
\newblock API:Main page - MediaWiki.
\newblock \url{https://www.mediawiki.org/wiki/API:Main_page}.
\newblock Accessed: 2025-01-15.

\bibitem[{Mehta(2024)}]{mehta_2024_social}
Mehta, I. 2024.
\newblock Social networks are getting stingy with their data, leaving third-party developers in the lurch | TechCrunch.
\newblock \url{https://techcrunch.com/2024/02/09/social-network-api-apps-twitter-reddit-threads-mastodon-bluesky/}.
\newblock Accessed: 2025-01-15.

\bibitem[{Meier(2022)}]{meier_twikiltwitter_2022}
Meier, F. 2022.
\newblock {TWikiL}–the {Twitter} {Wikipedia} {Link} {Dataset}.
\newblock In \emph{Proceedings of the {International} {AAAI} {Conference} on {Web} and {Social} {Media}}, volume~16, 1292--1301.

\bibitem[{Menking, Erickson, and Pratt(2019)}]{menking_people_2019}
Menking, A.; Erickson, I.; and Pratt, W. 2019.
\newblock People {Who} {Can} {Take} {It}: {How} {Women} {Wikipedians} {Negotiate} and {Navigate} {Safety}.
\newblock In \emph{Proceedings of the 2019 {CHI} {Conference} on {Human} {Factors} in {Computing} {Systems}}, {CHI} '19, 1--14. New York, NY, USA: Association for Computing Machinery.
\newblock ISBN 978-1-4503-5970-2.

\bibitem[{Miquel-Ribé and Laniado(2019)}]{miquel-ribe_wikipedia_2019}
Miquel-Ribé, M.; and Laniado, D. 2019.
\newblock Wikipedia cultural diversity dataset: {A} complete cartography for 300 language editions.
\newblock In \emph{Proceedings of the {International} {AAAI} {Conference} on {Web} and {Social} {Media}}, volume~13, 620--629.

\bibitem[{Mitrevski, Piccardi, and West(2020)}]{mitrevski_wikihist_2020}
Mitrevski, B.; Piccardi, T.; and West, R. 2020.
\newblock {WikiHist}. html: {English} {Wikipedia}'s full revision history in {HTML} format.
\newblock In \emph{Proceedings of the {International} {AAAI} {Conference} on {Web} and {Social} {Media}}, volume~14, 878--884.

\bibitem[{Moyer et~al.(2015)Moyer, Carson, Dye, Carson, and Goldbaum}]{moyer_determining_2015}
Moyer, D.; Carson, S.; Dye, T.; Carson, R.; and Goldbaum, D. 2015.
\newblock Determining the {Influence} of {Reddit} {Posts} on {Wikipedia} {Pageviews}.
\newblock \emph{Proceedings of the International AAAI Conference on Web and Social Media}, 9(5): 75--82.
\newblock Number: 5.

\bibitem[{Perez(2024)}]{perez_reddit_2024}
Perez, S. 2024.
\newblock Reddit locks down its public data in new content policy, says use now requires a contract.
\newblock \url{https://techcrunch.com/2024/05/09/reddit-locks-down-its-public-data-in-new-content-policy-says-use-now-requires-a-contract/}.
\newblock Accessed: 2025-01-14.

\bibitem[{Rascouët-Paz(2024)}]{rascouet-paz_elon_2024}
Rascouët-Paz, A. 2024.
\newblock Elon {Musk} {Urged} {People} to {Stop} {Donating} to {Wikipedia}. {Here}'s {Why}.
\newblock \url{https://www.snopes.com//fact-check/elon-musk-stop-donating-wikipedia/}.
\newblock Accessed: 2025-01-15.

\bibitem[{Reddit(2024)}]{_2024_developer}
Reddit. 2024.
\newblock Developer Terms.
\newblock \url{https://redditinc.com/policies/developer-terms}.
\newblock Accessed: 2025-01-15.

\bibitem[{Saez-Trumper(2019)}]{saez-trumper_online_2019}
Saez-Trumper, D. 2019.
\newblock Online {Disinformation} and the {Role} of {Wikipedia}.
\newblock \url{http://arxiv.org/abs/1910.12596}.
\newblock Accessed: 2025-01-13.

\bibitem[{Schulz, Fletcher, and Nielsen(2024)}]{schulz_role_2024}
Schulz, A.; Fletcher, R.; and Nielsen, R.~K. 2024.
\newblock The role of news media knowledge for how people use social media for news in five countries.
\newblock \emph{New Media \& Society}, 26(7): 4056--4077.
\newblock Publisher: SAGE Publications.

\bibitem[{Shen and Rosé(2022)}]{shen_tale_2022}
Shen, Q.; and Rosé, C.~P. 2022.
\newblock A {Tale} of {Two} {Subreddits}: {Measuring} the {Impacts} of {Quarantines} on {Political} {Engagement} on {Reddit}.
\newblock \emph{Proceedings of the International AAAI Conference on Web and Social Media}, 16: 932--943.

\bibitem[{Singer et~al.(2017)Singer, Lemmerich, West, Zia, Wulczyn, Strohmaier, and Leskovec}]{singer_why_2017}
Singer, P.; Lemmerich, F.; West, R.; Zia, L.; Wulczyn, E.; Strohmaier, M.; and Leskovec, J. 2017.
\newblock Why {We} {Read} {Wikipedia}.
\newblock In \emph{Proceedings of the 26th {International} {Conference} on {World} {Wide} {Web}}, {WWW} '17, 1591--1600.
\newblock ISBN 978-1-4503-4913-0.

\bibitem[{St.~Aubin and Liedke(2024)}]{st_aubin_news_2024}
St.~Aubin, C.; and Liedke, J. 2024.
\newblock News {Platform} {Fact} {Sheet}.
\newblock \url{https://www.pewresearch.org/journalism/fact-sheet/news-platform-fact-sheet/}.
\newblock Accessed: 2025-01-14.

\bibitem[{Taraborelli(2015)}]{taraborelli_sum_2015}
Taraborelli, D. 2015.
\newblock The sum of all human knowledge in the age of machines: a new research agenda for {Wikimedia}.
\newblock In \emph{{ICWSM}-15 {Workshop} on {Wikipedia}}.

\bibitem[{Tripodi(2023)}]{tripodi_ms_2023}
Tripodi, F. 2023.
\newblock Ms. {Categorized}: {Gender}, notability, and inequality on {Wikipedia}.
\newblock \emph{New Media \& Society}, 25(7): 1687--1707.

\bibitem[{Twyman, Keegan, and Shaw(2017)}]{twyman_black_2017}
Twyman, M.; Keegan, B.~C.; and Shaw, A. 2017.
\newblock Black {Lives} {Matter} in {Wikipedia}: {Collective} {Memory} and {Collaboration} around {Online} {Social} {Movements}.
\newblock In \emph{Proceedings of the 2017 {ACM} {Conference} on {Computer} {Supported} {Cooperative} {Work} and {Social} {Computing}}, {CSCW} '17, 1400--1412. Association for Computing Machinery.
\newblock ISBN 978-1-4503-4335-0.

\bibitem[{u/KeyserSosa(2024)}]{a2024_our}
u/KeyserSosa. 2024.
\newblock Our plans for Researchers on Reddit.
\newblock \url{https://www.reddit.com/r/reddit4researchers/comments/1co0mqa/our_plans_for_researchers_on_reddit/}.
\newblock Accessed: 2025-01-15.

\bibitem[{Valentim et~al.(2021)Valentim, Comarela, Park, and Sáez-Trumper}]{valentim_tracking_2021}
Valentim, R.~V.; Comarela, G.; Park, S.; and Sáez-Trumper, D. 2021.
\newblock Tracking knowledge propagation across wikipedia languages.
\newblock In \emph{Proceedings of the {International} {AAAI} {Conference} on {Web} and {Social} {Media}}, volume~15, 1046--1052.

\bibitem[{Vincent and Hecht(2021)}]{vincent_deeper_2021}
Vincent, N.; and Hecht, B. 2021.
\newblock A {Deeper} {Investigation} of the {Importance} of {Wikipedia} {Links} to {Search} {Engine} {Results}.
\newblock \emph{Proceedings of the ACM on Human-Computer Interaction}, 5(CSCW1): 1--15.

\bibitem[{Vincent, Johnson, and Hecht(2018)}]{vincent_examining_2018}
Vincent, N.; Johnson, I.; and Hecht, B. 2018.
\newblock Examining {Wikipedia} {With} a {Broader} {Lens}: {Quantifying} the {Value} of {Wikipedia}'s {Relationships} with {Other} {Large}-{Scale} {Online} {Communities}.
\newblock In \emph{Proceedings of the 2018 {CHI} {Conference} on {Human} {Factors} in {Computing} {Systems}}, 1--13. ACM.
\newblock ISBN 978-1-4503-5620-6.

\bibitem[{Vraga and Tully(2021)}]{vraga_news_2021}
Vraga, E.~K.; and Tully, M. 2021.
\newblock News literacy, social media behaviors, and skepticism toward information on social media.
\newblock \emph{Information, Communication \& Society}, 24(2): 150--166.

\bibitem[{{Wikimedia Foundation}(2024)}]{wikimedia_foundation_page_2024}
{Wikimedia Foundation}. 2024.
\newblock Page view.
\newblock \url{https://meta.wikimedia.org/wiki/Research:Page_view}.
\newblock Accessed: 2025-01-13.

\bibitem[{Wilkinson et~al.(2016)Wilkinson, Dumontier, Aalbersberg, Appleton, Axton, Baak, Blomberg, Boiten, da~Silva~Santos, and Bourne}]{wilkinson_fair_2016}
Wilkinson, M.~D.; Dumontier, M.; Aalbersberg, I.~J.; Appleton, G.; Axton, M.; Baak, A.; Blomberg, N.; Boiten, J.-W.; da~Silva~Santos, L.~B.; and Bourne, P.~E. 2016.
\newblock The {FAIR} {Guiding} {Principles} for scientific data management and stewardship.
\newblock \emph{Scientific data}, 3(1): 1--9.
\newblock Publisher: Nature Publishing Group.

\end{thebibliography}

\end{document}